# Breaking the 800 mV open-circuit voltage barrier in antimony sulfide photovoltaics

Jiacheng Zhou,[1] Xinwei Wang,[2] Tianle Shi,[1] Lei Wan,[1] Junzhi Ye,[3] Zhiqiang Li,[4] Aron Walsh,*,[2] Robert L. Z. Hoye,*,[3] and Ru Zhou*,[1]

[1] School of Electrical Engineering and Automation, Hefei University of Technology, Hefei 230009, P. R. China

[2] Department of Materials, Imperial College London, Exhibition Road, London SW7 2AZ, UK

[3] Inorganic Chemistry Laboratory, Department of Chemistry, University of Oxford, South Parks Road, Oxford OX1 3QR, United Kingdom

[4] National-Local Joint Engineering Laboratory of New Energy Photoelectric Devices, College of Physics Science and Technology, Hebei University, Baoding 071002, P. R. China

---

* Corresponding authors:

a.walsh@imperial.ac.uk (Aron Walsh)

robert.hoye@chem.ox.ac.uk (Robert L. Z. Hoye)

zhouru@hfut.edu.cn (Ru Zhou)

**Abstract**

$Sb_2S_3$ is a promising material for low-toxicity, high-stability next-generation photovoltaics. Despite high optical limits in efficiency, progress in improving its device performance has been limited by severe voltage losses. Recent spectroscopic investigations suggest that self-trapping occurs in $Sb_2S_3$, limiting the open-circuit voltage ($V_{OC}$) to a maximum of approximately 800 mV, which is the level the field has asymptotically approached. In this work, we surpass this voltage barrier through reductions in the defect density in $Sb_2S_3$ thin films by modulating the growth mechanism in chemical bath deposition using citrate ligand additives. Deep level transient spectroscopy identifies two deep traps 0.4–0.7 eV above the valence band maximum, and, through first-principles calculations, we identify these to likely be S vacancies, or Sb on S anti-sites. The concentrations of these traps are lowered by decreasing the grain boundary density from 1114±52 nm μm$^{-2}$ to 585±10 nm μm$^{-2}$, and we achieve a $V_{OC}$ of 824 mV, the record for $Sb_2S_3$ solar cells. This work addresses the debate in the field around whether $Sb_2S_3$ is limited by defects or self-trapping, showing that it is possible to improve the performance towards the radiative limit through careful defect engineering.

## Introduction

Thin film photovoltaics are important alternatives to silicon-based solar cells because of their high performance, lower materials use, compatibility with fabrication into larger-area modules, and lower levelized cost of electricity[1, 2]. As representatives, CdTe and Cu(In,Ga)(S,Se)$_2$ (CIGS) have achieved remarkable certified power conversion efficiencies (PCEs) of 23.1% and 23.6% at the laboratory scale, respectively[3]. Nevertheless, the use of toxic (i.e., Cd) and scarce (i.e., In, Ga) elements impede their widespread deployment on the multi-terawatt level. The ongoing development of highly efficient, cost-effective, earth-abundant solar cells that have low environmental impact and carbon dioxide equivalent (CO$_2$eq) footprint has led to the continuous emergence of light-harvesting materials for next-generation thin film photovoltaics, including Sb$_2$S$_3$, Sb$_2$Se$_3$, SnS, GeSe, CuSbS$_2$, CuSbSe$_2$, Cu$_2$ZnSnS$_4$, Cu$_2$ZnSn(S,Se)$_4$[4, 5, 6, 7, 8, 9, 10, 11, 12]. Antimony chalcogenides are at the forefront of emerging sustainable photovoltaic materials due to their excellent electronic and optical properties[4, 7, 13, 14, 15, 16]. This material system exhibits high absorption coefficient (>10$^5$ cm$^{-1}$ in the visible range), tunable bandgap (1.1-1.7 eV), decent charge-carrier mobility in thin films (~10 cm$^2$ V$^{-1}$ s$^{-1}$), and excellent thermal and environmental stability. Additionally, the quasi-one-dimensional (quasi-1D) crystal structure that comprises numerous [Sb$_4$S$_6$]$_n$ or [Sb$_4$Se$_6$]$_n$ ribbons enables efficient charge-carrier transport in the out-of-plane direction if the crystal orientation of antimony chalcogenide thin films can be effectively controlled[17, 18]. Antimony chalcogenide solar cells have recently exceeded a 10% efficiency benchmark based on Sb$_2$(S,Se)$_3$ absorbers, making them competitive in the emerging thin film photovoltaics field[5, 7, 13, 19].

The binary compound semiconductor Sb$_2$S$_3$ possesses a bandgap in the range of 1.7-1.8 eV, close to the optimal value for indoor photovoltaics (IPVs) and top-cells for silicon-based tandem photovoltaics[20, 21, 22]. According to the Shockley-Queisser (S-Q) detailed balance limit for the single p-n junction, under standard one-sun (AM1.5G, 100 mW cm$^{-2}$) illumination, the theoretical device parameters achievable with a 1.70 eV bandgap light-harvester are an open-circuit voltage ($V_{OC}$) of 1.402 V, a short-circuit current density ($J_{SC}$) of 22.46 mA cm$^{-2}$, a fill factor (FF) of 91%, and a power conversion efficiency (PCE) of 28.64%[23, 24]. During the past decade, Sb$_2$S$_3$ solar cells have made notable strides in performance. However, the most efficient Sb$_2$S$_3$ solar cells reported thus far (8.32% PCE) still falls far behind the detailed-balance limit[25]. Like other emerging light absorbers, the performance of Sb$_2$S$_3$ solar cells is primarily limited by the large $V_{OC}$ deficit [2]. At present, the $V_{OC}$ of Sb$_2$S$_3$ solar cells has stagnated in the range of 550-770 mV for decades, much smaller than other photovoltaic devices based on absorbers with similar or smaller bandgaps, such as CdTe, GaAs, CIGS, InP, CH$_3$NH$_3$PbI$_3$ or crystalline-Si, as summarized in Fig. 1a[1, 26]. To date, the $V_{OC}$ of 753 mV obtained in the highest-efficiency Sb$_2$S$_3$ device at present is only 53.7% of the radiative limit, suffering from a $V_{OC}$ deficit greater than 600 mV[25]. Here the $V_{OC}$ deficit is defined as $V_{OC}^{SQ} - V_{OC}$, where $V_{OC}^{SQ}$ is the S-Q limit $V_{OC}$, evaluated using the empirical equation $V_{OC}^{SQ} = 0.932 \times E_g/q - 0.167\ V$ ($E_g$ is the bandgap of absorber; Fig. 1a)[27]. Overcoming the $V_{OC}$ bottleneck is critical to ensure that progress with Sb$_2$S$_3$ photovoltaics does not stagnate.

The origin of the $V_{OC}$ bottleneck of Sb$_2$S$_3$ solar cells remains under debate, with two perspectives illustrated in Fig. 1b. One explanation proposed by Yang et al. is self-trapping of photoexcited charge-carriers, which sets an upper limit to the maximum $V_{OC}$ (~800 mV) and PCE (~16.00%) for Sb$_2$S$_3$ solar cells[28]. The proposed mechanism involves excitation from the ground state (GS) to excited state (ES), before relaxing to self-trapped exciton (STE) states due to electron-phonon coupling, leading to energy losses and reductions in the quasi-Fermi level splitting. This mechanism is linked to their experimental observation of a 600 meV red-shift in photoluminescence (PL) compared to the optical bandgap, along with polarized light emission from Sb$_2$S$_3$ single crystals. A similar carrier self-trapping effect has also been proposed for Sb$_2$Se$_3$[29]. If self-trapping were indeed limiting the performance of these materials, this places a fundamental limitation on the performance below the radiative limit, since carrier localization takes place intrinsically, even in a defect-free material. In contrast, first-principles analysis of electron-lattice interactions and charge-carrier mobility suggest that the low effective masses in $Sb_2S_3$ lead to band-like transport in these materials, and the key limiting factor for $V_{OC}$ is defect-mediated rather than self-trapping[30]. Several experimental studies have linked the high $V_{OC}$ deficit in $Sb_2S_3$ solar cells to a complex set of defects with deep levels[27, 31, 32]. For example, we previously showed that lowering the defect density in $Sb_2S_3$ by reducing the grain boundary (GB) density results in an increase in device performance, including improvements in $V_{OC}$[14]. In-depth investigations combining theory and experiments are needed to resolve this debate to understand whether $Sb_2S_3$ is fundamentally limited in performance, or has hope still of further improvements towards the radiative limit. But beyond mechanistic studies, it is essential to demonstrate if devices surpassing the limits expected from self-trapping could be realized.

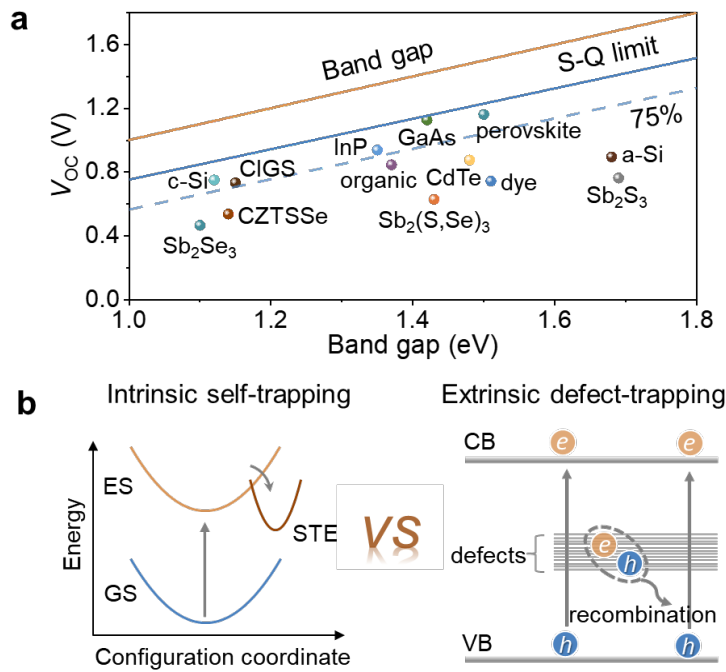

**Fig. 1 | Proposed mechanisms for factors limiting the open-circuit voltage of $Sb_2S_3$ solar cells. a** Reported $V_{OC}$ values of well-developed thin film solar cells, as well as the most efficient antimony chalcogenides[1, 26]. Theoretical S-Q limit of $V_{OC}$ as a function of band gap (solid blue line) and 75% of the limit (dashed line) are highlighted. **b** Qualitative adiabatic potential energy curve (left) showing the self-trapping effect: photoexcitation from ground state (GS) to excited state (ES) evolves into a self-trapped exciton (STE) state from lattice distortions and losing substantial energy.

Schematic (right) showing the defect-trapping effect: the photoexcitated electrons and holes suffer from charge recombination at the defects.

In this work, we developed $Sb_2S_3$ solar cells with the $V_{OC}$ breaking the 800 mV barrier by using a complexing agent-mediated chemical bath deposition (CBD) method, which enables the deposition of high-quality $Sb_2S_3$ thin films with GB density approximately halved. The reduced impact of defects for $Sb_2S_3$ films, confirmed by deep-level transient spectroscopy (DLTS) characterization and first-principles calculations, is revealed to contribute to the smaller $V_{OC}$ deficit. The obtained $V_{OC}$ of 824 mV is the current record for antimony chalcogenide solar cells. The trap-limited $V_{OC}$ and PCE as well as the radiative limit of $Sb_2S_3$ solar cells was also assessed by computations. This work reveals that defect-mediated recombination is a key limitation to the device $V_{OC}$ and PCE, and the low performance of $Sb_2S_3$ photovoltaics is surmountable through defect engineering to reach towards the radiative limit.

## Results

### Performance of planar $Sb_2S_3$ solar cells

In this work, a modified CBD method was explored to deposit large-grained $Sb_2S_3$ absorber films. It is known that additive engineering in CBD solutions by incorporating reasonable complexing agents is an effective strategy to regulate the film properties (such as surface morphology, GB density, defect density, carrier concentration, etc.) via the facile modulation of *in-situ* chemical environment for film deposition[23, 33]. Here, potassium antimony tartrate ($C_4H_4KO_7Sb·0.5H_2O$, APT) and sodium thiosulfate ($Na_2S_2O_3·5H_2O$, STS) were employed as the Sb source and S source, respectively, and a small amount of sodium citrate (SC) was used as a complexing additive in the precursor solution to manipulate the nucleation and growth kinetics of $Sb_2S_3$ films. For the convenience of description, the $Sb_2S_3$ samples obtained with the SC concentrations of 1 mM, 2 mM, 4 mM, and 8 mM in the precursor solutions are labeled as 1mM-SC, 2mM-SC, 4mM-SC, and 8mM-SC, respectively. The planar n-i-p device configuration used is FTO/SnO$_2$/CdS/Sb$_2$S$_3$/Spiro-OMeTAD/Au (Fig. 2a). Here Sb$_2$S$_3$ is the absorber, and SnO$_2$/CdS and Spiro-OMeTAD act as the electron transport layer (ETL) and hole transport layer (HTL), respectively. The detailed fabrication processes of Sb$_2$S$_3$ solar cells are illustrated in Supplementary Fig. 1. The photovoltaic parameters of Sb$_2$S$_3$ solar cells, each based on 12 cells, including $V_{OC}$, $J_{SC}$, FF, and PCE, achieved under one-sun illumination (AM1.5G, 100 mW cm$^{-2}$), are summarized in the box charts of Fig. 2b. The corresponding device parameters are summarized in Table 1. As shown, these devices exhibit excellent performance and repeatability. The device PCE increases at first and then decreases with the increase of the SC concentration in the CBD precursor solution. The control Sb$_2$S$_3$ devices obtained without the use of SC deliver an average PCE of 6.23%, whereas the average PCE is increased to 7.46% for the 4mM-SC devices. The current density-voltage (*J-V*) curves of best-performing control and 4mM-SC Sb$_2$S$_3$ solar cells are given in Fig. 2c. The control device yields a $V_{OC}$ of 787 mV and a PCE of 6.42%; the 4mM-SC device affords a $V_{OC}$ of 817 mV and a PCE of 7.67%. That is, the incorporation of 4mM SC leads to the relative PCE increasing by nearly 20% compared to the control device. The external quantum efficiency (EQE) spectra of the control and 4mM-SC solar cells, as given in Fig. 2d, reveal that both devices present broad light responses ranging from 350 to 750 nm. Compared to the control device, the 4mM-SC device exhibits enhanced EQE, with the value exceeding 90% at around 530 nm. The obtained integrated $J_{SC}$ from EQE spectra are 14.95 and 16.13 mA cm$^{-2}$ for the control and 4mM-SC devices, respectively, consistent with that obtained from *J-V* curves (within 5% deviation). Fig. 2e summarizes the $V_{OC}$ evolution of well-developed planar and sensitized Sb$_2$S$_3$ solar cells (detailed in Supplementary Table 1). As shown, the $V_{OC}$ presents continuous enhancements over the past decade, especially for the planar configuration. However, the $V_{OC}$ shows relatively slow growth in contrast to the rapid PCE increase. To the best of our knowledge, the $V_{OC}$ of 824 mM obtained for the 4mM-SC device is the highest value reported for Sb$_2$S$_3$ solar cells. Interestingly, this $V_{OC}$ value is beyond the predicted limit for Sb$_2$S$_3$ if it had self-trapping[28]. We also found that CBD-processed Sb$_2$S$_3$ photovoltaics exhibited excellent environmental stability. The unencapsulated 4mM-SC Sb$_2$S$_3$ solar cell retained 97.4% of its initial efficiency after 30 days of storage in the dark under ambient temperature in a cabinet with 15% relative humidity (Supplementary Fig. 2).

Moreover, it is known that the bandgap of $Sb_2S_3$ is ideal for IPVs[33]. Thus we further measured the IPV performance of $Sb_2S_3$ solar cells under illumination from white light emitting diodes (WLEDs) with a correlated color temperature (CCT) of 3000 K, and details of the spectral irradiance, the stability of light source, and indoor photovoltaic device performance for 4 mM-SC are shown in Supplementary Fig. 3 and 4, along with Supplementary Table 2. The short-circuit current density is approximately linearly proportional to the light intensity, while the $V_{OC}$ and FF increase approximately logarithmically with the light intensity. As a result, the highest $Sb_2S_3$ IPV efficiency is under 1000 lux, with a state-of-the-art indoor PCE of 18.56%. This exceeds our previous report of $Sb_2S_3$ IPVs (17.6% indoor PCE), and is comparable with the current state-of-the-art for not just $Sb_2S_3$, but also other Sb- and Bi-based solar absorbers[33, 34, 35]. The loss in $V_{OC}$ from 1000 lux to 200 lux in the 4 mM-SC IPVs is 61 mV, which is low compared to other established and emerging IPVs[36].

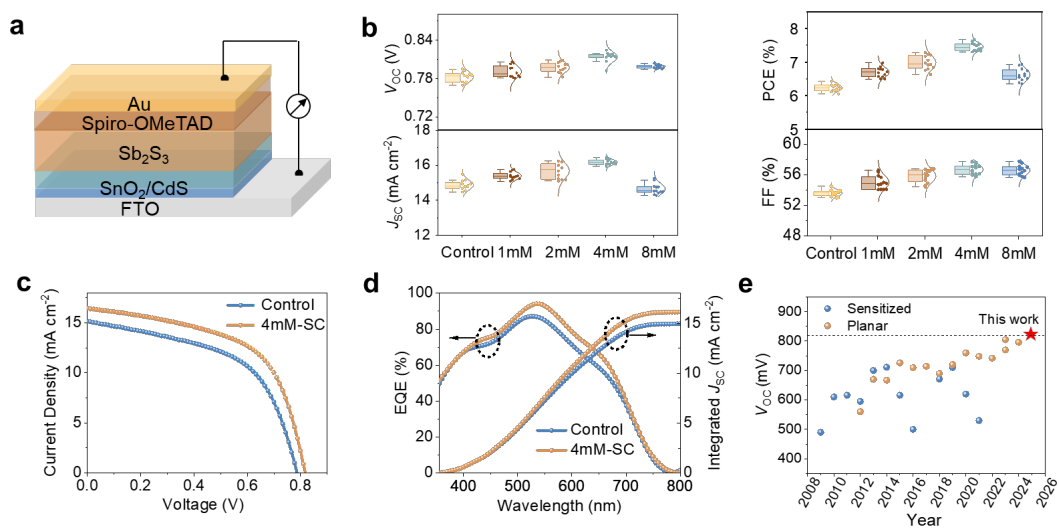

**Fig. 2 | Photovoltaic performance of planar $Sb_2S_3$ thin film solar cells. a** The planar configuration of FTO/SnO$_2$/CdS/Sb$_2$S$_3$/Spiro-OMeTAD/Au used in this work. **b** Statistics of the performance parameters of $Sb_2S_3$ solar cells obtained without (i.e., control) and with the addition of different SC concentrations. **c** Current density-voltage (*J-V*) curves of best-performing control and 4mM-SC

Sb$_2$S$_3$ solar cells, measured under one-sun (AM 1.5G, 100 mW cm$^{-2}$) illumination. **d** External quantum efficiency (EQE) spectra of control and 4mM-SC Sb$_2$S$_3$ solar cells. **e** Evolution in the record $V_{OC}$ values of planar and sensitized Sb$_2$S$_3$ solar cells.

Table 1 | Photovoltaic parameters of the control and SC-Sb$_2$S$_3$ solar cells, measured under one-sun (AM 1.5G, 100 mW cm$^{-2}$) illumination. Format: average ± standard deviation (best value)

| Device | $V_{OC}$ (mV) | $J_{SC}$ (mA cm$^{-2}$) | FF (%) | PCE (%) |
|---|---|---|---|---|
| Control | 782±8 | 14.87±0.2 | 53.60±0.42 | 6.23±0.11 |
|  | (795) | (15.15) | (53.90) | (6.42) |
| 1mM-SC | 791±10 | 15.40±0.21 | 55.02±0.90 | 6.70±0.15 |
|  | (806) | (15.76) | (56.58) | (6.98) |
| 2mM-SC | 797±8 | 15.69±0.45 | 55.85±0.86 | 6.98±0.21 |
|  | (807) | (16.23) | (56.75) | (7.29) |
| 4mM-SC | 814±8 | 16.18±0.15 | 56.68±0.67 | 7.46±0.12 |
|  | (824) | (16.44) | (57.70) | (7.67) |
| 8mM-SC | 799±3 | 14.65±0.31 | 56.60±0.67 | 6.62±0.19 |
|  | (803) | (15.26) | (56.99) | (6.93) |

**Characterization of CBD-processed Sb$_2$S$_3$ films**

Fig. 3a and b shows top-view scanning electron microscopy (SEM) images of Sb$_2$S$_3$ thin films prepared without the SC additive (i.e., the control sample) and with different concentrations of 4 mM SC, respectively. It can be seen that, the control Sb$_2$S$_3$ film sample displays grain sizes smaller than 3 μm based on the long diameters of polygon-shaped grains, consistent with that reported for CBD-processed Sb$_2$S$_3$ films in the literature[4]. With the addition of SC, the grain size of Sb$_2$S$_3$ films involves an obvious increase, with many Sb$_2$S$_3$ grains exceeding 5 μm. Top-view SEM images of Sb$_2$S$_3$ films prepared with the use of other SC concentrations are provided in Supplementary Fig. 5. The increase in grain size is accompanied by a decrease in GB density, which is defined as the GB length per unit area. Fig. 3a gives the dependence of GB density on the SC concentration for different $Sb_2S_3$ films. As shown, the GB density on the surface of $Sb_2S_3$ films decreases from 1114±52 nm μm$^{-2}$ for the control sample to 666±17, 641±16, 586±11 and 901±31 nm μm$^{-2}$ for 1mM-SC, 2mM-SC, 4mM-SC and 8mM-SC film samples, respectively. In polycrystalline films, the recombination losses are largely caused by GBs where dangling or wrong bonds proliferate[14]. Hence, the performance could be increased for polycrystalline absorbers with reduced GB density.

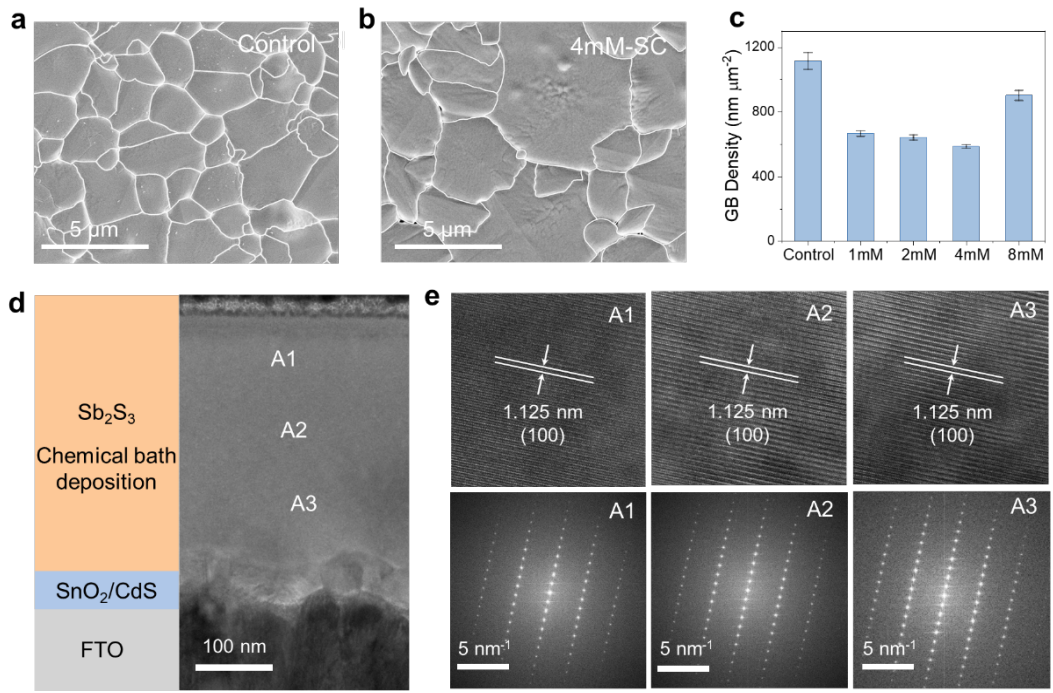

**Fig. 3 | Morphological properties of highly crystalline $Sb_2S_3$ thin films. a,b** Top-view scanning electron microscopy (SEM) images of $Sb_2S_3$ films prepared without (i.e., control) and with the addition of 4 mM SC. The GBs are highlighted to clearly show the changes in GB density. **c** Change in GB density with SC concentration in the precursor solution. Three samples were used to determine the mean GB density values shown, and the error bars denote the standard deviation. **d** Cross-sectional transmission electron microscopy (TEM) image of 4mM-SC $Sb_2S_3$ sample deposited on FTO/$SnO_2$/CdS substrate. **e** High-resolution TEM images performed at points A1, A2, and A3 (see part D), and corresponding fast Fourier transform (FFT) patterns.

To further confirm the single-crystalline nature of individual $Sb_2S_3$ grains observed from SEM images, we performed characterization based on the $Sb_2S_3$ lamella sample (4mM-SC film) prepared by focused ion beam (FIB). The cross-sectional transmission electron microscopy (TEM) image (Fig. 3b) indicates that the $Sb_2S_3$ sample is compact and composed of large grains spanning the thickness of the $Sb_2S_3$ film. Fig. 3c shows the high-resolution TEM (HRTEM) images of three arbitrarily selected points A1, A2, and A3, as labeled in Fig. 3b. As shown, points A1, A2, and A3 share identical crystallographic orientations, revealing that the crystal planes extend from the top to the bottom of the $Sb_2S_3$ film. Lattice fringes with interplanar spacings of 1.125 nm, corresponding to the (100) plane in orthorhombic $Sb_2S_3$, were identified[14]. The nearly identical fast Fourier transform (FFT) patterns of HRTEM images of points A1, A2, and A3 further confirm that the individual microstructural units bounded by GBs are indeed single crystals. The cross-sectional elemental mapping images (Supplementary Fig. 6) indicate the uniform distribution of Sb and S across the absorber layer and the conformal growth of $Sb_2S_3$ on top of the CdS buffer layer. The HRTEM image at the $CdS/Sb_2S_3$ interface (Supplementary Fig. 7) reveal the lattice fringes assigned to $Sb_2S_3$ and CdS[14]. The results confirmed that $Sb_2S_3$ films with large single-crystalline grains could be easily produced via CBD.

X-ray diffraction (XRD) patterns of $Sb_2S_3$ films (Supplementary Fig. 8a) show that the diffraction peaks can be well indexed to orthorhombic $Sb_2S_3$ (JCPDS #42-1393)[23]. The diffraction peaks of $Sb_2S_3$ films reveal negligible shifts with the use of SC, indicating that the SC additive could not be incorporated into the $Sb_2S_3$ lattice or interstitial sites. We further calculated the texture coefficient (TC) of the dominant ($hk$0) and ($hk$1) planes to quantify the degree of preferred orientation of $Sb_2S_3$ films (Supplementary Fig. 8b). The addition of SC does not cause significant changes to the film orientation. Additionally, compared to the control film, the $SC-Sb_2S_3$ films involve a gradual reduction in the dominant diffraction peak intensities, such as (120), (130), (211), (221), with the increase in the SC concentration. This should be closely associated with the decrease in the film thickness, as reflected in the cross-sectional SEM images of $Sb_2S_3$ films (Supplementary Fig. 9). As shown, the film thickness gradually decreases from 506 ± 26 nm for the control sample to 394 ± 10 nm, 347 ± 28 nm, 318 ± 15 nm and 116 ± 32 nm for 1mM-SC, 2mM-SC, 4mM-SC, and 8mM-SC samples, respectively.

The two-dimensional (2D) morphology spatial maps of atomic force microscopy (AFM) characterization (Supplementary Fig. 10) confirm the 4mM-SC film shows larger grain sizes with fewer GBs than the control. Furthermore, the addition of SC leads to a reduction in the root-mean-square roughness from 21.9 nm (the control sample) to 16.3 nm (the 4mM-SC sample). The reduced surface roughness of the absorber films would favor the formation of an improved heterojunction between $Sb_2S_3$ and solution-processed HTL on top. Conductive-atomic force microscopy (c-AFM) images, with the corresponding morphology and current maps (Supplementary Fig. 11) reveal that, compared to the control sample, the 4mM-SC films have more uniform and slightly enhanced electrical conductivity over the 5 × 5 μm$^2$ area measured. The reduced fluctuation of the current intensity for the 4mM-SC $Sb_2S_3$ film should be conducive to the local photocurrent generation and collection, which might effectively suppress charge-carrier recombination at GBs[37]. Moreover, the GBs of $Sb_2S_3$ grains generally involve lower conductivity than grain interiors, suggesting the inefficient carrier separation and transport derived from defects at the GBs[38]. We further performed Kelvin probe force microscopy (KPFM) to map the surface potential of $Sb_2S_3$ films (Supplementary Fig. 12), which reveals a higher surface potential for the 4mM-SC film in contrast to the control sample. Regarding that an increase in the surface potential corresponds to a reduced work function (WF) (i.e., closer to the vacuum level), the smaller WF for the 4mM-SC sample implies the upshift of Fermi levels[37, 39]. This is consistent with the band structure of $Sb_2S_3$ solar cells, obtained from ultraviolet photoelectron spectroscopy (UPS) spectra (Supplementary Fig. 13) and absorption spectra (Supplementary Fig. 14). Compared to the control $Sb_2S_3$ film, the Fermi level of SC-$Sb_2S_3$ is shifted up towards conduction band (CB), indicating an increase in the electron carrier concentration for n-type $Sb_2S_3$.

We further investigate the underlying mechanism that could explain the impact of the SC complexing agent on $Sb_2S_3$ film properties. For the CBD of $Sb_2S_3$ films, the substrate is exposed to soluble precursors which react or decompose to form the desired film materials[40]. From the perspective of nucleation and growth, an intricate balance of growth parameters is necessary to achieve high-quality films. The formation of $Sb_2S_3$ grains typically involves nucleation and growth, which are greatly influenced by the chemical environment of precursor solutions[41, 42, 43]. We measured the contact angles of the substrate treated by precursor solutions with different SC concentrations and further analyzed the ratio of the free energies of homogeneous and heterogeneous nucleation as a function of the contact angle (Supplementary Fig. 15a, b, detailed in Supplementary Note 1). The results demonstrate that the incorporation of SC would promote the heterogeneous nucleation of $Sb_2S_3$ on the substrate. Our proposed mechanism is schematically illustrated in Supplementary Fig. 15c. We suggest that adding SC changes the growth mode of $Sb_2S_3$ films from cluster deposition to ion-by-ion deposition. For the pristine CBD solution without the use of SC, the strong complexation effect of ATP would cause the formation of many aggregates in the precursor, which serve as homogeneous nucleation centers for $Sb_2S_3$[44]. These small clusters created within the solution further migrate and adhere onto the substrate surface, resulting in the deposition of $Sb_2S_3$ films in the cluster mechanism. For the ion-by-ion deposition mechanism, the complexing agent enables the slow release of free metal ions to directly react with anions onto the substrate in a more controlled manner, being capable of forming well-textured crystalline or even monocrystalline films[40]. Ion-by-ion deposition therefore promotes heterogeneous nucleation on the substrate and suppresses homogeneous nucleation in the precursor solution, similar to the scenario demonstrated for the CBD of other metal chalcogenide thin films, such as ZnS[45]. High-density small nuclei formed on the substrate would further coalesce into a large-grained $Sb_2S_3$ film during the CBD process. Hence the formation of stable citrate-complexed $Sb^{3+}$ succeeds to control the nucleation and growth rate of $Sb_2S_3$, contributing to the deposition of large-grained $Sb_2S_3$ films. In fact, previous studies also demonstrated that films grown by ion-by-ion deposition are usually associated with larger crystalline size compared to those grown by cluster deposition[40]. Additionally, the decrease in the release rate of $Sb^{3+}$ further reduces the film thickness, explaining the thickness evolution of $Sb_2S_3$ films with the addition of SC. Apart from the complexation of citrate, we speculate that pH might be another critical factor influencing the deposition of $Sb_2S_3$ films. The influence of pH on the deposition of $Sb_2S_3$ films is further discussed in Supplementary Note 2. It is revealed that, besides the complexation of $Sb^{3+}$ with citrate, the increase in pH should also contribute to the reduction in the deposition rate. However, controlling pH alone does not lead to increases in PCE as high as with the addition of SC (Supplementary Fig. 16). The results imply that the increase in PCE for SC-$Sb_2S_3$ solar cells should be mainly associated with the complexation $Sb^{3+}$ with citrate which causes the evolution in the morphology of $Sb_2S_3$ absorber films.

**Defect physics and charge-carrier transport properties**

Understanding the nature of active defects in $Sb_2S_3$ is necessary to design strategies to minimize their impact. DLTS is a powerful technique offering insights into the trap energy, defect type and defect concentration in thin film solar cells[46, 47, 48, 49]. DLTS utilizes the transient capacitance of a p-n junction at varying temperatures as a probe to reflect the changes in the charge state of a deep defect center. The traps within the device are filled with carriers by applying a voltage pulse, which in turn alters the capacitance associated with the p-n junction. Fig. 4a, b illustrate the mechanism of DLTS measurements. Here we take the minority traps (i.e., electron traps in p-type semiconductors) as an example to show the change of transient capacitance caused by electron capture and emission. For a p-n junction, when a constant reverse bias ($U_R$) is applied, a stable space charge region (SCR) is established, corresponding to a stable capacitance ($C_R$). If an extremely short fill voltage pulse ($U_P$) is applied as a perturbation, the SCR width and capacitance change due to the injection of majority carriers into the SCR, causing the traps to be filled during this process. Once the voltage pulse is removed, the SCR and capacitance cannot immediately recover to their original state. This delayed recovery occurs because the trapped carriers are emitted to the CB or valence band (VB) at a specific rate, rather than instantaneously. Consequently, a transient capacitance change over time is observed, and the dynamic capacitance can be described using the Eq. 1[49]:

$$C(t) = C_R + \Delta C * \exp\left(-\frac{t}{\tau_e}\right) \qquad (1)$$

where $\Delta C$ is the amplitude of the transient capacitance and $\tau_e$ is the emission time constant, which is the reciprocal of emission rate, expressed by the Eq. 2[50]:

$$\tau_e = \frac{t_2 - t_1}{\ln\left(\frac{t_2}{t_1}\right)} \qquad (2)$$

As $\Delta C$ and $\tau_e$ vary with temperature, a sequence of the transient capacitance is further collected at various temperatures. The capacitance changes within a time window ($\Delta C = C_{t2} - C_{t1}$) vs. the different temperatures ($T$) are sampled as the DLTS signals. For the analysis of DLTS measurements, the activation energy ($E_a$) and capture cross-section ($\sigma$) of defects are determined from the Arrhenius plots using Eqs. 3 and 4 6:

$$\ln(\tau_e v_{\text{th,n}} N_C) = \frac{E_C - E_T}{k_B T} - \ln(X_n \sigma_n) \qquad (3)$$

$$\ln(\tau_e v_{\text{th,p}} N_V) = \frac{E_T - E_V}{k_B T} - \ln(X_p \sigma_p) \qquad (4)$$

where $\tau_e$ denotes the emission time constant, $E_C$, $E_V$, and $E_T$ are the CB minimum (CBM), valance band maximum (VBM), and defect energy level, respectively. $v_{\text{th,n}}$ and $v_{\text{th,p}}$ indicate the thermal velocities for electron and hole traps, respectively. $X_n$ and $X_p$ are the entropy factor for hole and electron, respectively. $N_C$ and $N_V$ represent the density of states of CB and VB, respectively. $\sigma_n$ and $\sigma_p$ represent the capture cross-section of electron and hole traps, respectively. $T$ and $k_B$ are the temperature and Boltzmann constant, respectively. $E_a$ is calculated using $E_C - E_T$ for electron trap and $E_T - E_V$ for hole trap. $E_a$ can be obtained from the slope of Arrhenius plots through linear regression. $\sigma$ can be extracted from the intersection of a line with the $y$-axis. The trap density ($N_T$) can be obtained using Eq. 5[49]:

$$N_T = 2N_S \Delta C / C_R \qquad (5)$$

where $N_S$ represents the shallow donor/acceptor concentration in absorber films, $C_R$ and $\Delta C$ represent the capacitance at reverse bias in equilibrium and the amplitude of capacitance transient, respectively. In particular, the trap type is determined by the $\Delta C$. If $\Delta C < 0$ and the DLTS signal peaks are negative, they correspond to the minority-carrier trap; in contrast, they are majority-carrier traps[49, 50, 51].

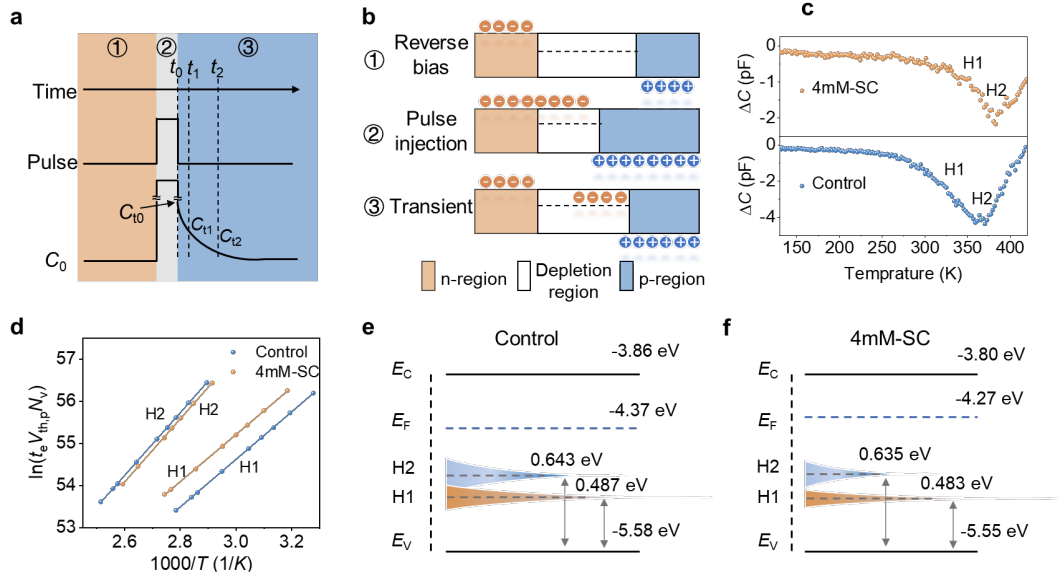

**Fig. 4 | DLTS characterization of defect physics in $Sb_2S_3$ solar cells. a** Schematic illustration of the mechanism of DLTS measurement. Here we take the minority traps (i.e., electron traps in p-type semiconductors) as an example to illustrate the change of transient capacitance caused by electron capture and emission. **b** The variation of depletion width and the process of carriers being trapped and emitted during the measurement. **c** DLTS signals from the control and 4mM-SC $Sb_2S_3$ solar cells. **d** Arrhenius plots derived from DLTS signals. Data points are achieved by calculating internal transients included in DLTS signals with the discrete Laplace transform, and the solid lines are corresponding linear fits. **e,f** Schematic of band edge positions and defect levels of the control and 4mM-SC $Sb_2S_3$, respectively, including CBM ($E_C$) and VBM ($E_V$), Fermi level ($E_F$) and defect energy levels (H1, H2), relative to the vacuum level.

**Table 2 | Defect characteristic parameters, including trap energy level ($E_T$), cross section ($\sigma$), and trap concentration ($N_T$), obtained using DLTS characterization in the control and 4mM-SC $Sb_2S_3$ solar cells.**

| Devices | Trap | $E_T$ [eV] | $\sigma$ [cm$^2$] | $N_T$ [cm$^{-3}$] | $\sigma \times N_T$ [cm$^{-1}$] |
|---|---|---|---|---|---|
| Control | H1 | $E_V$+0.487 | 4.43×10$^{-17}$ | 2.42×10$^{15}$ | 0.107 |
|  | H2 | $E_V$+0.643 | 7.35×10$^{-16}$ | 1.39×10$^{15}$ | 1.020 |
| 4mM-SC | H1 | $E_V$+0.483 | 2.08×10$^{-17}$ | 9.14×10$^{14}$ | 0.019 |
|  | H2 | $E_V$+0.635 | 6.62×10$^{-16}$ | 1.03×10$^{15}$ | 0.682 |

As shown in Fig. 4c, the control device and 4mM- device both present two negative DLTS signal peaks in the high temperature region, denoted as H1 and H2, due to the effect of multiple minority-carrier traps, i.e., hole defects in n-type $Sb_2S_3$ films[6]. The defect characteristic parameters, including trap energy ($E_T$) and capture cross section ($\sigma$), and trap concentration ($N_T$), can be extracted from the linear fittings of the Arrhenius plots as given in Fig. 4d, with the results summarized in Table 2. Both devices share similar types of deep levels with densities of approximately $10^{15}$ cm$^{-3}$. The trapping behavior can be quantified by using the thermal trapping rate ($C_{trap}$), expressed by Eq. 6[48]:

$$C_{trap} = 1/\tau_{trap} = v\sigma N_T \tag{6}$$

where $\tau_{trap}$ is the carrier lifetime associated with the trap-assisted Shockley-Read-Hall (SRH) recombination, which is inversely proportional to $\sigma \times N_T$, and $v$ is the thermal velocity of charge related to the intrinsic charge transport feature of semiconductors. The values of $\sigma \times N_T$ for the control and 4mM-SC devices are summarized in Table 2. Compared to H1, H2 dominates the charge trapping in absorber films with a larger $\sigma \times N_T$ value and greater energy depth. Compared to the control device, the $\sigma \times N_T$ value for the H2 defect in 4mM-SC $Sb_2S_3$ is decreased. The band edge position and defect levels are illustrated in Fig. 4e, f, with the hole traps situated 0.4-0.7 eV above the VBM. Since the trap energy reflects the difficulty of charge emission from the traps, the H1 and H2 defects located at more than 0.3 eV above the VB would trap carriers without re-emission and act as recombination centers due to their high ionization energy, contributing to severe non-radiative recombination. In particular, the deep-level defects would cause a fluctuation in the bandgap and bring band tail states within the bandgap, giving rise to a further $V_{OC}$ loss. Hence the 4mM-SC $Sb_2S_3$ film with reduced defect density is suggested to effectively suppress charge recombination in solar cells.

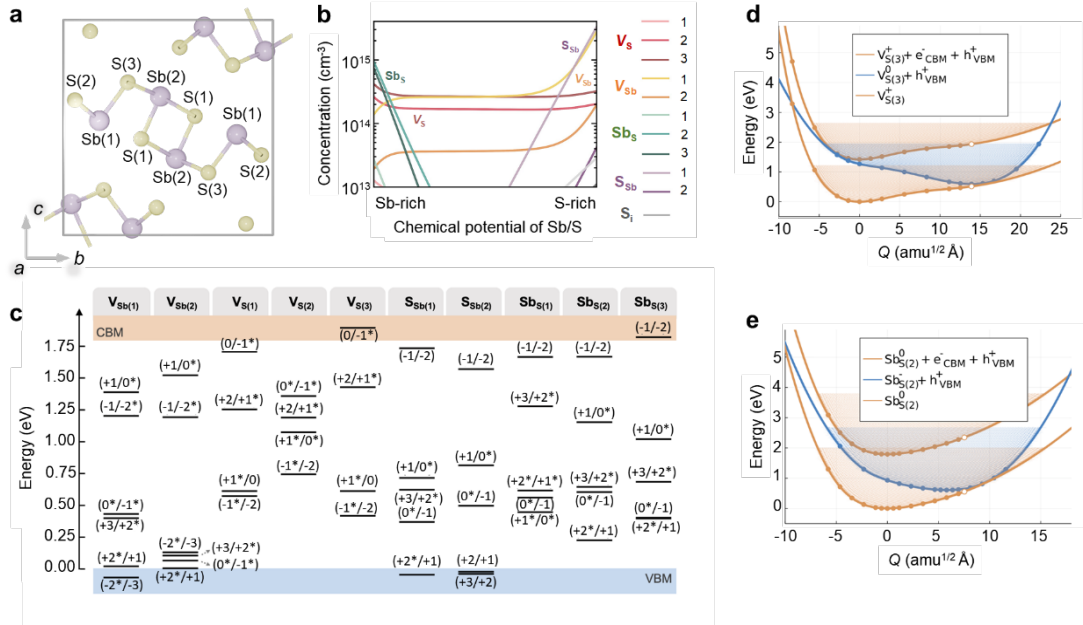

**Fig. 5 | First-principles calculations to identify defect species. a** Illustration of a $[Sb_4S_6]_n$ unit in the $Sb_2S_3$ crystal structure. **b** Calculated equilibrium defect concentration at 300 K in $Sb_2S_3$ crystals grown at 643 K as a function of growth conditions based on the defect energies reported in Ref.[52]. The numbers in the legend represent different inequivalent sites. **c** Calculated charge transition levels of intrinsic point defects with high concentrations under Sb-rich conditions in $Sb_2S_3$. Metastable charge states are marked with asterisks (*). The Fermi level is referenced to the valence band maximum (VBM). Dashed grey lines indicate trap levels in control and 4mM-SC $Sb_2S_3$ devices, determined by DLTS. **d,e** 1D configuration coordinate diagram for charge transitions between $V_{S(3)}^+$ and $V_{S(3)}^0$ as well as $Sb_{S(2)}^0$ and $Sb_{S(2)}^-$. $Q$ represents the generalized configuration coordinate, which describes the collective atomic displacement along the reaction pathway during charge transition. The equilibrium structures of $V_{S(3)}^+$ and $Sb_{S(2)}$ are set as references with $Q = 0$ amu$^{1/2}$Å and total energy $E = 0$ eV. Solid circles are data points obtained by DFT calculations and used for fitting, while hollow circles are discarded for fitting due to charge delocalization. Solid lines represent best fits to the data.

To identify the defect species, we performed first-principles calculations combined with a global searching strategy, implemented in the doped[53] and ShakeNBreak packages[54], to systematically explore all symmetry-inequivalent configurations of intrinsic point defects in $Sb_2S_3$. It is known that quasi-1D $Sb_2S_3$ consists of $[Sb_4S_6]_n$ chains which have five nonequivalent atomic sites, i.e., Sb(1), Sb(2), S(1), S(2) and S(3), as illustrated in Fig. 5a. First principles calculations have shown that $Sb_2S_3$ involves intricate defect types, including five vacancies ($V_{Sb(1)}$, $V_{Sb(2)}$, $V_{S(1)}$, $V_{S(2)}$ and $V_{S(3)}$), five anti-sites ($S_{Sb(1)}$, $S_{Sb(2)}$, $Sb_{S(1)}$, $Sb_{S(2)}$ and $Sb_{S(3)}$), and twelve interstitials. The interstitials were identified using the Voronoi scheme[55], despite the material being comprised of only two elemental components[52, 56, 57]. Our calculation details are provided in Ref.[52] and the Methods section. The equilibrium defect concentrations as a function of the Sb/S chemical potential are shown in Fig. 5b. Under Sb-rich conditions (i.e., synthesis conditions for control and 4mM-SC $Sb_2S_3$ devices), $Sb_S$, $V_S$ and $V_{Sb}$ are the most dominant point defects with concentrations higher than $10^{14}$ cm$^{-3}$. The corresponding defect levels are shown in Fig. 5c and summarized in Supplementary Table 3. Except for $V_{Sb(2)}$, most of these defects have traps that match those observed experimentally via DLTS (as indicated by the dashed gray lines). This highlights the challenges in identifying specific defects based solely on defect concentrations or energy levels in $Sb_2S_3$. Moreover, our global search reveals that all intrinsic point defects in $Sb_2S_3$ are amphoteric, exhibiting both donor- and acceptor-like charge states. Therefore, directly comparing experimental defect levels with simulations under the assumption that each defect acts exclusively as a donor or an acceptor may be misleading[56], since amphoteric defects introduce additional possibilities for matching the measured levels.

While defect formation energies and defect levels calculated from simulations provide insights into the thermodynamic stability, they do not directly account for kinetic factors such as carrier capture processes under real device conditions. Experimental techniques are sensitive to defects that actively capture carriers. To identify the most relevant defects, we first selected those with defect levels and equilibrium concentrations close to the experimental values (Supplementary Table 4), then investigated the potential energy surfaces associated with charge transition processes. Capture cross sections were then calculated for each transition using configuration coordinate (CC) diagrams[52]. The magnitude of the cross sections was used to determine whether each defect predominantly behaves as an electron or a hole trap. Based on the combined information from defect levels, concentrations, and cross sections, $V_{S(3)}$ and $Sb_{S(2)}$ were identified as the defects most consistent with the experimental observations, with trap levels located at 0.59 and 0.61 eV above the VBM and calculated capture cross sections of $3.77 \times 10^{-16}$ and $4.00 \times 10^{-17}$ cm$^{-2}$, respectively. However, due to the small differences between these trap levels, it is challenging to unambiguously assign $V_{S(3)}$ and $Sb_{S(2)}$ to H1 and H2. One possible reason for the discrepancy between theoretical predictions and experimental observations is the absence of temperature effects in our electronic structure calculations. In real device conditions, defect occupation probabilities and nonradiative recombination dynamics could be influenced by thermal activation. Such temperature effects can lead to a shift of 0.1-0.2 eV in defect levels, partly accounting for the observed differences[58]. The CC diagrams for $V_{S(3)}$ and $Sb_{S(2)}$ are shown in Fig. 5d, e. A larger displacement along the generalized coordinate $Q$ corresponds to stronger lattice relaxation during the charge transition process. For both defects, the energy barriers for hole capture are very small, and the corresponding phonon wavefunctions of the initial and final states exhibit large overlap. A small capture barrier combined with large phonon overlap greatly enhances the transition probability, leading to large hole capture cross-sections on the order of $10^{-16} \sim 10^{-17}$ cm$^2$.

We further predicted the maximum achievable $V_{OC}$ and PCE by accounting for both radiative (band-to-band) and defect-mediated non-radiative recombination processes (Supplementary Fig. 17). A film thickness of 350 nm and optical absorption coefficients simulated from first-principles calculations[49] (HSE06 functional with a bandgap of 1.79 eV), were used as inputs to ensure a consistent radiative limit for both samples. Under the radiative limit, the calculated $V_{OC}$ is 1.70 V, with a corresponding maximum PCE of 21.98%. Non-radiative recombination losses were assessed based on measured trap levels, trap densities and capture cross sections in control and 4mM-SC $Sb_2S_3$ devices. Since only hole traps are detected by DLTS, we assume non-radiative recombination in these samples is dominated by hole capture processes (i.e., electron capture coefficients are set infinitely high). For the control sample, the predicted $V_{OC}$ is 1.15 V, corresponding to a PCE of 13.34% and a voltage loss $\Delta V_{OC}$ of 0.35 V (defined as $V_{OC}^{SQ} - V_{OC}$, where $V_{OC}^{SQ}$ is calculated to be around 1.50 V). In contrast, the 4mM-SC sample shows an improved $V_{OC}$ of 1.26 V and a PCE of 14.72%, with a reduced $\Delta V_{OC}$ of 0.24 V. These predictions agree well with the observed enhanced performance of the 4mM-SC $Sb_2S_3$ device, but both samples exhibit significant non-radiative losses compared to the radiative limit, highlighting the role of defect-mediated recombination.

**Device physics of planar $Sb_2S_3$ solar cells**

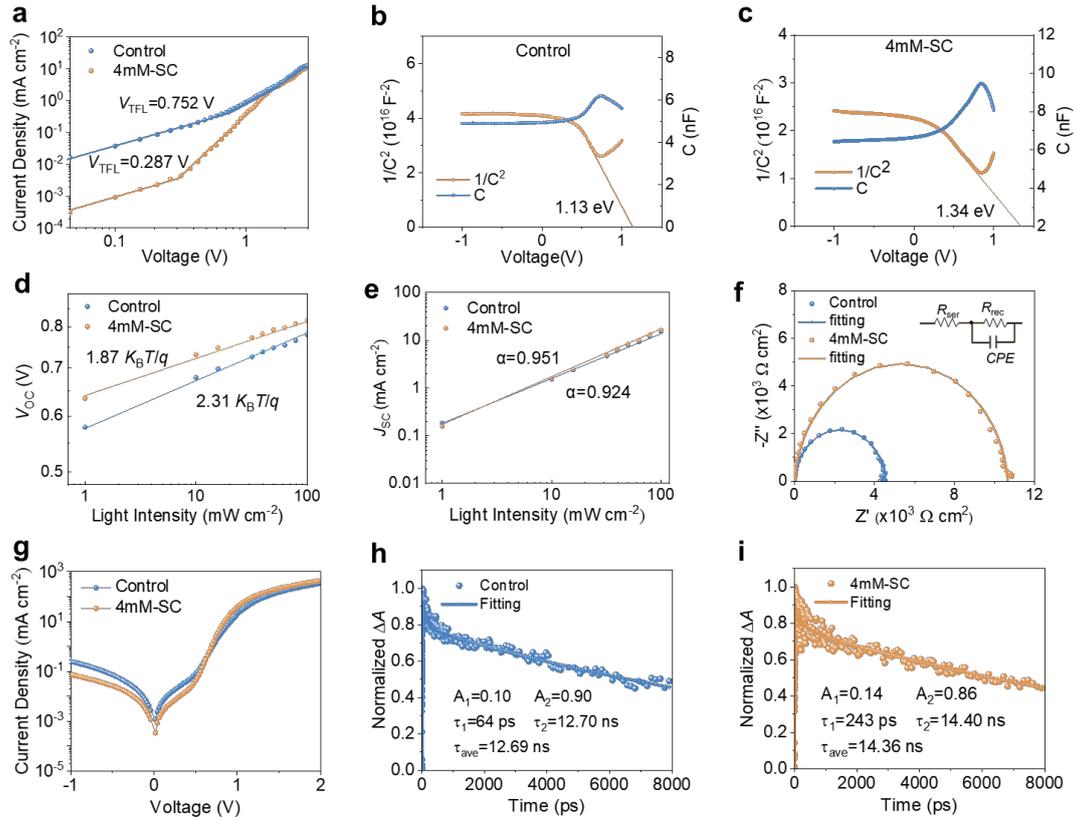

**Fig. 6 | Device physics of planar $Sb_2S_3$ solar cells. a** Space charge limited current (SCLC) measurements of the control and 4mM-SC devices based on the electron-only device with the configuration of FTO/CdS/$Sb_2S_3$/PCBM/Au. **b,c** Capacitance-voltage (*C-V*) and corresponding Mott-Schottky ($1/C^2$~*V*) plots of the control and 4mM-SC $Sb_2S_3$ solar cells measured in the dark. **d,e** Dependences of $V_{OC}$ and $J_{SC}$ on the light intensity for the control and 4mM-SC $Sb_2S_3$ solar cells. **f** Nyquist plots for the control and 4mM-SC devices measured in the dark, with the equivalent circuit used for fitting shown in the inset. **g** *J-V* curves measured in the dark for the control and 4mM-SC $Sb_2S_3$ solar cells. **h,i** Photoinduced absorption (PIA) kinetics (scatter) measurements monitored at 560 nm from transient absorption spectroscopy, along with phenomenological biexponential curve

fits (solid line) for the of the control and 4mM-SC $Sb_2S_3$ films. Excitation was with a 400 nm wavelength pulsed laser at a repetition rate of 1000 Hz. $\Delta A$ is defined as the change in the absorption of the sample before and after pumping.

We further investigate the device physics of $Sb_2S_3$ solar cells to fully understand underlying mechanism of performance enhancements. The space charge limited current density (SCLC) measurements by the dark logarithmic *J-V* curves of the electron-only device FTO/CdS/$Sb_2S_3$/PCBM/Au is given in Fig. 6a. At the low voltage Ohmic region, the curve is generally linear. When the voltage exceeds the trap-filled limiting region, the current increases dramatically, implying that the injected carriers have completely occupied the trap states[25]. The trap-filled limiting voltage ($V_{TFL}$) values for the control and 4mM-SC device are estimated to be 0.287 V and 0.752 V, respectively. The trap state density ($n_\tau$) of $Sb_2S_3$ films can be evaluated according to the equation $V_{TFL} = qn_\tau L^2/2\varepsilon_r\varepsilon_0$, where $q$ is the elementary charge, $n_\tau$ is the trap state density, $L$ is the thickness of the absorber film (here 506 nm and 318 nm for the control and 4mM-SC $Sb_2S_3$ films, respectively), $\varepsilon_0$ is the vacuum permittivity (8.85 × $10^{-12}$ F $m^{-1}$), and $\varepsilon_r$ is the relative permittivity of $Sb_2S_3$ (6.67 for static dielectric constant)[59]. As a result, the trap state densities for the control and 4mM-SC device are estimated to be 2.17×$10^{15}$ $cm^{-3}$ and 2.12×$10^{15}$ $cm^{-3}$, respectively. The overall reduction in trap state density is consistent with the reduction in GB density, and the order of magnitude of the trap density is consistent with DLTS measurements.

Fig. 6b, c shows the capacitance-voltage (*C-V*) and corresponding Mott-Schottky ($1/C^2$~*V*) plots of the control and 4mM-SC $Sb_2S_3$ devices measured in the dark. It is demonstrated that the abrupt p-n junction capacitance can be expressed using a parallel-plate capacitor model, in which the relationship between the capacitance and the bias voltage is described as $1/C^2 = 2(V_{bi} - V)/q\varepsilon_0\varepsilon_r A^2 N_{CV}$[27], where $q$ is the elementary charge (1.60×$10^{-19}$ C), $\varepsilon_0$ is the vacuum permittivity (8.85×$10^{-12}$ F $m^{-1}$), $\varepsilon_r$ is the relative permittivity of $Sb_2S_3$ (6.67), $A$ is the active area of solar cells (0.06 $cm^2$), $V_{bi}$ is the built-in potential, and $N_{CV}$ is the charge-carrier density. According to this equation, the $V_{bi}$ of the 4mM-SC device is calculated to be 1.34 V, larger than that of the control device

(1.13 V). The increase in the $V_{bi}$ reflects the enhanced band bending at the heterointerface, which facilitates the charge separation and reduces the interface recombination. A larger $V_{bi}$ of the depletion region contributes to the enhanced $V_{OC}$, consistent with the results from J-V curves.

The dependences of $V_{OC}$ and $J_{SC}$ on the light intensity for both solar cells were performed to reveal trap-assisted charge recombination loss mechanisms, as shown in Fig. 6d, e. The relationship between $V_{OC}$ (and $J_{SC}$) and the light intensity $I$ can be described by $V_{OC} \propto (nk_BT/q)\ln(I)$ and $J_{SC} \propto I^{\alpha}$, where $k_B$ is the Boltzmann constant, $T$ is the absolute temperature, $q$ is the elementary charge, and $\alpha$ and $n$ reflect the level of charge-carrier recombination[60, 61, 62]. Under the open-circuit condition, all photogenerated charge-carriers recombine. For ideal trap-free solar cells, the slope of $V_{OC}$ versus $\ln(I)$ is $k_BT/q$. As revealed, the slopes for the control and 4mM-SC devices are linearly fitted to be $2.31k_BT/q$ and $1.87k_BT/q$, respectively. The reduced slope for the 4mM-SC device reflects the suppression of trap-assisted charge-carrier recombination. The parameter $\alpha$ ideally equals 1, and $\alpha < 1$ suggests the loss of photogenerated charge-carriers caused by the incomplete collection. The $\alpha$ values for the control and 4mM-SC devices are estimated to be 0.924 and 0.951, respectively. The increase in $\alpha$ for the 4mM-SC device implies the enhancement of the carrier collection rate[62]. Impedance spectroscopy analysis, Fig. 6f and Supplementary Table 5, shows the 4mM-SC device delivers an increased recombination resistance at the CdS/Sb$_2$S$_3$ interface ($R_{rec}$=10.66 kΩ cm$^{-2}$) compared to the control device ($R_{rec}$=4.51 kΩ cm$^{-2}$). The increase in the resistance is suggested to suppress the charge recombination and improve the charge collection in solar cells.

Fig. 6g presents J-V curves of the control and 4mM-SC devices measured under dark conditions. According to the abrupt junction J-V equation and its formula manipulation, the parameters of junction ideality factor ($A$), the reverse saturation current density ($J_0$), the series resistance ($R_S$), and the shunt conductance ($G$, i.e., $1/R_{SH}$)) can be extracted (Supplementary Fig. 18)[61]. The detailed analysis is given in Supplementary Note 3. It can be seen that, compared to the control device, the 4mM-SC device has lower $R_S$ and $G$, which implies an improvement in the carrier extraction capability. The *A* values extracted from the control and 4mM-SC devices are 2.38 and 1.91, respectively; the reduced *A* value indicates the suppression of defect states in the absorber films[61]. The calculated $J_0$ is reduced from $3.22\times10^{-5}$ mA cm$^{-2}$ for the control device to $2.09\times10^{-6}$ mA cm$^{-2}$ for the 4mM-SC device. The main reason for the generation of $J_0$ is the charge-carrier recombination caused by deep-level defects in the devices, and thus the reduced $J_0$ implies the suppression of the defect-induced recombination, contributing to the $V_{OC}$ increase for the 4mM-SC device[63].

Ultrafast transient absorption (TA) spectroscopy analysis was further performed to understand the charge-carrier kinetics of FTO/SnO$_2$/CdS/Sb$_2$S$_3$ films. The TA spectra of both the control sample and the 4mM-SC sample (Supplementary Fig. 19a, b) display distinct negative differential absorbance (ground state bleach (GSB) peaks), as well as regions of positive differential absorbance (photoinduced absorption (PIA) peaks). The GSB peaks observed at the wavelengths of 460-520 nm and 610-680 nm can be ascribed to the state filling of CdS and the ground state absorption of Sb$_2$S$_3$, respectively. The PIA peak at the wavelength of 520 to 610 nm can be assigned to the formation of sulfide radicals ($S^{-\bullet}$) as a result of the localization of photogenerated holes on the S atom within the Sb$_2$S$_3$ lattice[6, 14, 64]. The transient kinetic decays monitored at 560 nm for both the control and 4mM-SC films can be extracted from the pseudocolor images for the TA spectra of both samples (Supplementary Fig. 19c, d), as presented in Fig. 6h, i. The transient dynamics are well fitted with a phenomenological biexponential model in order to quantitatively compare between samples. The gradual decrease of the PIA peak can be attributed to the decay of trapped holes, i.e., the $S^{-\bullet}$ species, which we here attribute to the nonradiative carrier recombination in Sb$_2$S$_3$ films. The 4mM-SC sample delivers higher average time constant ($\tau_{av}$) values (6.61 ns) compared to the control sample (4.93 ns). Such lifetime values lies within the range of 5~60 ns that are commonly reported for antimony chalcogenides[31]. In general, the charge-carrier lifetime of the absorber material is inversely proportional to reverse saturation current density of the device. Hence, a long carrier lifetime is the prerequisite for achieving high $V_{OC}$ of solar cells.

**Discussion**

While it has been suggested that the low $V_{OC}$ of Sb$_2$S$_3$ solar cells is an intrinsic limit caused by self-trapping in the bulk material, we have succeeded in overcoming the predicted 800 mV barrier. This achievement was realized through the formation of large-grained thin films by adding SC to the CBD precursor solution. The growth of high-quality uniform films produces Sb$_2$S$_3$ solar cells with PCE of 7.67% under one-sun illumination and a record $V_{OC}$ of 824 mV. Defect characterization and computations reveal that the reduced defect activity in Sb$_2$S$_3$ films contribute to the increase in device $V_{OC}$. We conclude that antimony chalcogenides have the potential to be used in high-performance optoelectronic devices. While intrinsic defects do give rise to non-radiative recombination of photogenerated charge carriers, there is plenty of scope to further increase the performance of this technology through effective defect engineering and crystal growth strategies.

**Acknowledgements**

J.Z., X.W. and T.S. contributed equally to this work. The financial support by the National Natural Science Foundation of China (nos. 52371219 and W2421064) and the Fundamental Research Funds for the Central Universities (no. JZ2024HGTG0295) is greatly acknowledged. J. Y. and R. L. Z. H. thank UK Research and Innovation for funding through a Frontier Grant (no. EP/X029900/1), awarded via the 2021 ERC Starting Grant scheme. They also thank St. John's College Oxford for funding through the Large Grant scheme. R. L. Z. H. thanks the Royal Academy of Engineering and Science & Technology Facilities Council for funding through a Senior Research Fellowship (no. RCSRF2324-18-68). R. Z. and R. L. Z. H. thank the Royal Society for funding (no. IEC/NSFC/242353).

## Author contributions

R.Z. and J.Z. designed the experiments and analyzed the data, and also conceived of the idea for this manuscript. J.Z. and T.S. carried out the experiments, device optimizations and data analysis. X.W. carried out the computations. L.W., J.Y., Z.L., A.W., R.L.Z.H. and R.Z. assisted in experiments and data analysis. R.Z., J.Z., X.W., A.W. and R.L.Z.H., wrote the manuscript. All authors commented on the manuscript.

## Conflict of interest

The authors declare no competing interests.

## Supplementary Materials

The online version contains supplementary materials available at xxx